# Developing Structural, High-heat flux and Plasma Facing Materials for a near-term DEMO Fusion Power Plant: the EU Assessment.

D Stork[a,1§], P Agostini[b], J-L Boutard[c], D Buckthorpe[d], E Diegele[e], S L Dudarev[a], C English[f], G Federici[g], M R Gilbert[a], S Gonzalez[g], A Ibarra[h], Ch Linsmeier[i], A Li Puma[j], G Marbach[c], P F Morris[k,2], L W Packer[a], B Raj[l], M Rieth[e], M Q Tran[m], D J Ward[a] and S J Zinkle[n]

[a] *Euratom - CCFE Association, Culham Science Centre, Abingdon, Oxfordshire OX14 3DB, UK*
[b] *ENEA, Brasimone Research Centre, 40032, Cumugnano, Bologna, Italy*
[c] *CEA, cab HC, Saclay, F-91191, Gif-sur-Yvette, France*
[d] *AMEC, Booths Park, Chelford Road, Knutsford, Cheshire, WA16 8QZ, UK.*
[e] *Karlsruhe Institute for Technology, IMF-I, D-7602, Karlsruhe,, Germany*
[f] *National Nuclear Laboratory, Chadwick House, Warrington Road, Birchwood Park, WA3 6AE, UK*
[g] *EFDA Power Plant Physics and Technology, Boltzmannstr. 2, Garching, 85748 Germany*
[h] *CIEMAT, Avda. Complutense 40, Madrid,, Spain*
[i] *Forschungszentrum Jülich GmbH, Institut für Energie- und Klimaforschung – Plasmaphysik, EURATOM Association, 52425 Jülich, Germany*
[j] *CEA, DEN, Saclay, DM2S, SERMA, F-91191 Gif-sur-Yvette, France*
[k] *formerly of TATA steel Europe, Swinden Technology Centre, Moorgate, Rotherham S60 3AR, UK*
[l] *Indian National Academy of Engineering, Shaheed Jeet Singh Marg, New Delhi 110016, India*
[m] *Ecole Polytechnique Federale de Lausanne – CRPP, Association Euratom-Switzerland, 1015 Lausanne. Switzerland,*
[n] *Oak Ridge National Laboratory, P.O. Box 2008, Oak Ridge, TN, 37831, USA*
[1] *Present address: techne-physis Limited, Rivendell, The Causeway, Steventon, Oxon OX13 6SJ, UK*
[2] *Present address: 43 Hallam Road, Rotherham S60 3ED, UK*

*Abstract:*

The findings of the EU 'Materials Assessment Group' (MAG), within the 2012 EU Fusion Roadmap exercise, are discussed. MAG analysed the technological readiness of structural, plasma facing and high heat flux materials for a DEMO concept to be constructed in the early 2030s, proposing a coherent strategy for R&D up to a DEMO construction decision.

A DEMO phase I with a 'Starter Blanket' and 'Starter Divertor' is foreseen: the blanket being capable of withstanding $\geq 2$ MW.yr.m$^{-2}$ fusion neutron fluence (~20 dpa in the front-wall steel). A second phase ensues for DEMO with $\geq 5$ MW.yr.m$^{-2}$ first wall neutron fluence. Technical consequences for the materials required and the development, testing and modelling programmes, are analysed using: a systems engineering approach, considering reactor operational cycles, efficient maintenance and inspection requirements, and interaction with functional materials/coolants; and a project-based risk analysis, with R&D to mitigate risks from material shortcomings including development of specific risk mitigation materials. The DEMO balance of plant constrains the blanket and divertor coolants to remain unchanged between the two phases. The blanket coolant choices (He gas or pressurised water) put technical constraints on the blanket steels, either to have high strength at higher temperatures than current baseline variants (above 650ºC for high thermodynamic efficiency from He-gas coolant), or superior radiation-embrittlement properties at lower temperatures (~290-320ºC), for construction of water-cooled blankets. Risk mitigation proposed would develop these options in parallel, and computational and modelling techniques to shorten the cycle-time of new steel development will be important to achieve tight R&D timescales. The superior power handling of a water-cooled divertor target suggests a substructure temperature operating window (~200-350ºC) that could be realised, as a baseline-concept, using tungsten on a copper-alloy substructure. The difficulty of establishing design codes for brittle tungsten puts great urgency on the development of a range of advanced ductile or strengthened tungsten and copper compounds.

---

[§] Corresponding author. Address as 1. email: derek.stork@btinternet.com

Lessons learned from Fission reactor material development have been included, especially in safety and licensing, fabrication/joining techniques and designing for in-vessel inspection. The technical basis of using the ITER licensing experience to refine the issues in nuclear testing of materials is discussed.

Testing with 14MeV neutrons is essential to Fusion Materials development, and the Roadmap requires acquisition of ≥30 dpa (steels) 14MeV test data by 2026. The value and limits of pre-screening testing with fission neutrons on isotopically- or chemically-doped steels and with ion-beams are evaluated to help determine the minimum 14 MeV testing programme requirements.

# 1. Introduction

Successful operation of a 'Demonstration Reactor' (DEMO), capable of generating net electricity and operating with a closed fuel-cycle, is a key step in the development of fusion power. The recent EU Fusion Roadmap [1] has proposed a focussed R&D programme aimed at readiness to construct a 'near-term' DEMO, and highlights, *inter alia*, the 'Materials Mission' to develop nuclear-hardened structural, plasma facing and high-heat flux materials for DEMO's in-vessel components, especially the Breeding Blankets and Divertor. As part of the Roadmap exercise a 'Materials Assessment Group' (MAG) examined the technological readiness of these materials, identifying the major knowledge gaps and risks; and recommending a coherent strategy and R&D road map up to a decision point to construct DEMO in the early 2030s. The MAG has emphasised a systems-engineering approach: considering whole-system reactor operational cycles; needs for efficient maintenance and inspection; and interaction with functional materials/coolants. This has been supplemented by lessons to be learned from Fission reactor material development especially in safety and licensing, fabrication/joining techniques, designing for in-vessel inspection and development of manufacturing and supply-chain. The overall approach of treating Material Development as a project and analysing risks to the DEMO mission from knowledge gaps and known shortcomings has been taken. Risk mitigation measures are then proposed, including, where necessary the accelerated development of 'risk mitigation materials'. The MAG's scope of reference was limited to structural materials and plasma facing and high-heat flux materials, but interfaces with functional materials were also considered where necessary. The MAG findings are contained in a report [2], and have been incorporated into the Roadmap document. This paper reviews the

MAG methodologies, key technical findings and proposed development plans.

## 2. Characteristics of the EU Roadmap DEMO concept

The Roadmap foresees a decision to construct at DEMO machine in the early 2030s, following the important results from the ITER D-T programme and particularly the ITER Test Blanket Module (TBM) programme. This near-term DEMO is envisaged as a long-pulse (several hours) device, with the realistic extrapolation from ITER parameters consistent with the goal of ~500MW of net electricity generation to the grid and a fully-closed tritium fuel cycle. The requirements for DEMO are integrated into a coherent 0-D systems code [3], which provides a reference concept for the Roadmap and Materials assessments. The resulting near-term DEMO concept is described in more detail in [4], but its main characteristics are listed below:

| | | | | |
|---|---|---|---|---|
| | Major radius: | 9m | Minor radius: | 3.0m |
| , | Plasma current: | 20.3MA | Toroidal field | 5.3T |
| | Fusion power | 1.8$GW_{th}$ | Heating&Current Drive power | 50MW |
| | Pulse length | ~2.5hrs (with current drive assist) | | |

Plasma normalized pressure $\beta_N$ ~ 2.57 %MA.m$^{-1}$.T$^{-1}$.

The *baseline* DEMO concept has a water-cooled divertor and a helium-cooled Breeding Blanket. This baseline choice is discussed in more detail below. It is intended to be reviewed at the Conceptual Design Review of the EU DEMO in 2020, possibly to lead to a change of baseline, with considerations on materials being a key input to the decision.

### 2.1 Component testing philosophy

The EU DEMO is proposed to have two phases of test programme operation. In 'Phase I' the breeding blanket and divertor concepts will be refined with the possibility of installing modified versions in the subsequent 'Phase II', where the final, reactor, concepts will be proven. Thus between the phases it is envisaged that the technologies of the Breeding Blanket (BB) and divertor might change, including possibly the structural and functional materials. The only constraint imposed on the changes would be that the *coolants* for the BB and divertor could not be changed between phases. This constraint arises from the need to avoid changing the *Balance of Plant* (BoP), as such a change would be extremely difficult to achieve, unacceptably high in cost, both in capital and, equally importantly in a 'fast track' approach, in delays to completion of the programme. The Remote Handling technology and principles to be used in the change-out of the in-vessel components are the object of extensive current study in the European programme [5].

To define the DEMO Phase I programme, 'starter' blanket and divertor concepts have been developed. The DEMO starter BB should have a first wall structure capable of withstanding $\geq 2$ MW.yr.m$^{-2}$ fusion neutron fluence (equivalent to ~20 dpa front-wall steel damage). The starter divertor materials should match this in terms of their own neutron flux at the front face, and withstand the accompanying erosion damage from the plasma's scrape-off layer particle flux. For Phase II, the BB first wall structure should withstand $\geq 50$ dpa steel damage. The target machine availabilities (needed to determine the testing phases' durations) are 30% for Phase I and 50% for Phase II.

3. **Essential methods in the MAG Assessment**

   **3.1 Project-oriented risk analysis**

   A *Project-based approach* to the development of materials was adopted, with evaluation of risks and mitigating strategies, and linkage to decision points in the concept, system and detailed design processes of the relevant in-vessel systems.

   **3.2 Systems Engineering approach**

Additionally a *systems engineering approach*, similar to that of the EU DEMO design process [4], was adopted to: analyse the problems of sub-systems with assemblies of materials; highlight trade-offs and constraints; and prioritise R&D needs. The MAG has used the concept of Technology Readiness Levels (TRLs) applied in an approximate manner to the relevant materials. This has already been applied to fusion materials elsewhere [6].

**3.3 Lessons from Fission Material development**

The DEMO in-vessel environment will be markedly different from those in fission reactors, but there are key aspects from fission materials developments that may contribute to defining a DEMO R&D materials programme and provide some experience to optimise development. In particular, fission programmes include development of materials used in an irradiation environment where mechanical and dimensional properties suffer significant degradation. The implications of fission practice to materials R&D issues relevant to the design and licensing of DEMO and the lessons learnt were examined in the MAG process, and are detailed in section 3.5 below.

**3.4 Main elements of the Risk Analysis – Baseline and Risk Mitigation materials**

The process of the Risk Analysis and its interaction with the proposed R&D programme is shown in figure 1. For each of the materials categories (Structural, HHF and PF) a survey was made of current R&D knowledge, including materials development *per se*, joining and fabrication technologies, mechanical and thermo-mechanical properties, interaction with relevant coolants, and design code status. The irradiation loadings and the likely mechanical and thermal loadings were taken from the concept simulations. Irradiation effects were considered, and also the performance requirements for the in-vessel systems linkages to other DEMO sub-systems, especially BoP. Thus *Baseline Material(s)* was (were) identified for the immediate pursuance of the DEMO concept design. These materials were selected as the best available which had progressed at least beyond TRL3 (roughly 'proof of concept', with basic properties and performance known over the relevant operational range, joining techniques investigated, coolant compatibility demonstrated and irradiation effects measured), and had at least some of the elements of TRL4-5 demonstrated (roughly 'relevant multi-effects in integrated environments', with joining techniques demonstrated, large quantity fabrication, prototypes built and operated in simulated integrated environment, data and models existing on irradiation and coolant corrosion effects, and relevant nuclear use codes in existence) [6].

For the Baseline Material a *Risk Log* was developed listing risks to the DEMO mission from the known limitations, or unknown properties of the material. Risks were categorized from 'Low' to 'Very High', using the usual definition of Risk Level = Impact x Likelihood. For each risk above 'Moderate' level an identified *Mitigation Strategy,* was used, in the form of a constraint on the design, an operational 'work-around', more basic R&D, or the development of a new material. As shown in figure 1, post-mitigation risks were evaluated assuming a successful process. If use of the Baseline Material still entailed risks above 'Moderate', the development of substitute materials, categorized as *Risk Mitigation Materials (RMM)*, was included in the R&D process. RMMs included in the development programme were restricted to those existing at least at TRL1, ie. there is existing applied research on the material in basic form, and existing conceptual studies of the material's application. The MAG then fixed on the known *Issues* for particular RMMs, and established *'Issues Tables'* for RMMs which had progressed to the TRL3 stage. Addressing these issues, when added to the other risk mitigation measures aimed at improving the baseline, then formed the proposed R&D programme for a material category. As shown in figure 1, the RMMs and the improved Baseline would be developed in parallel, until appropriate down-selection points were reached, based on the stage of the DEMO in-vessel components' design. In down-selection, the RMM could replace the Baseline Material if the latter remained too risky, or the RMM represented a quantifiable improvement. After these selection points the balance of R&D resources would thus be altered. A non-selected RMM could still continue in development under the Roadmap's 'Advanced Concepts' programme [1], at a lower resource level, and without the near-term DEMO project time constraints.

For the HHF and PF categories of materials, current RMMs are at a very early development stage (TRL1-2). As such, it was not possible to establish accurate Issues Tables, so for these RMMs a generic materials development programme was identified, based on timely achievement of the TRL4-5 level.

**3.5 Key lessons from Fission programmes**

Fission experience shows that to develop and select materials for plants and components operating in irradiated environments the most important issues are:

   - the existence of suitable and commercially available materials;

   - the existence of a robust supply-chain for component design and manufacture;

   - the prior development and qualification of joining techniques, especially welding, which meet the rigorous scrutiny of regulators and operators;

   - the availability at an early stage of design codes to cover engineering design, including repair and replacement (equivalent to a 'Data Handbook');

   - an experimentally-validated assessment of degradation mechanisms and failure probabilities (including stress-corrosion cracking and coolant interface effects), backed up with accurate operational conditions modelling and experimental validation;

- in particular, regarding the lesson mentioned above, one should cite the importance of having available temperature-controlled irradiation test facilities which reproduce accurately the correct neutron spectrum and flux range characteristics of the power core system under development. This has proved effective in the development of Light Water Reactors (LWR) (for, eg. accurate end-of-life predictions and residual element limits in Reactor Pressure Vessel steels)[7, 8], and the Fast Breeder Reactors (FBR)(for, eg. fuel cladding and wrapper materials development)[9].

To achieve these requirements, it is estimated for the 'Generation IV' advanced fission systems, a programme of 10-15 years duration [10], is required to take a material from TRL1 to the threshold of reactor deployment. Thus any fusion materials required for an early-2030s DEMO decision should essentially already be at least at TRL1-2 level, and completely new developments are ruled out on this timescale.

*3.5.1 Regulatory factors in materials development*

Notwithstanding the comments in section 5.1 about structural materials for the ITER blanket, there are no established fusion design codes or practices to assure nuclear safety and a consistent engineering approach, but fission experience shows that greatest weight will be given to real plant experience and precedence. For ITER the French licensing authorities have accepted Safety Case proposals as the essential safety arguments have been met. These are based largely on the extension of established methodologies such as passive-safety containment and defence in depth, developed within the nuclear industry [11]. The regulatory perspectives of a DEMO safety case should be established at an early stage, and should include as much use of ITER precedent as possible.

The 'Fusion Materials Data Handbook' available in the early stages of the design

process will not include data for true 'fusion neutron spectrum' irradiations (see section 7). The situation is not so different from that encountered in the early phases of the fission reactor programme, and there the various programmes handled the issues by a combination of in-service monitoring of degradation and the use of mechanistic 'degradation formulae' (for example the prediction of in-service and end-of-life properties of Reactor Pressure Vessels (RPVs) in Fission Light Water Reactors (LWRs)). The RPV is the main safety-related component in an LWR, representing the primary containment barrier against Design Basis Accidents (DBAs), and is thus equivalent to the Vacuum Vessel in ITER. For the early LWRs, standard irradiation degradation factors were allowed in design substantiation [eg. 12], but tests showed that embrittlement would be much higher, thus surveillance programmes began to monitor the change of RPV properties [eg.13]. Such fission experience will probably prohibit use of similar formulaic degradation factors at the DEMO Licensing Stage. In fission systems the levels of damage at which *un-irradiated* material properties can be used in design and substantiation are very low in Safety Related (SR) components, for example US Nuclear Regulations require the effect of radiation to be detailed for RPV components experiencing a dose of $> 10^{17}$ n.cm$^{-2}$ at E>1MeV or >0.00015 dpa. The equivalent in a fusion reactor has to be regarded as very low *high-energy* (above most transmutation reactions' energy threshold) accumulated neutron fluence at the primary safety boundary. For ITER, the Vacuum Vessel (VV) has been established as the Primary Safety Barrier, as it is a natural boundary and can be easily connected to expansion volumes. As will be discussed in section 7, neutronics simulations show that the neutron flux to which it is subjected is very much reduced ( by more than two orders of magnitude) and exhibits a much softer spectrum (negligible fluence above 1 MeV), than at the in-vessel BB first wall. The MAG recommends that ITER practice in this Safety Case respect is carried over to DEMO, with Safety Analysis to establish the primary containment boundary for DEMO at an early stage in design.

*3.5.2 In-service Inspection and Maintenance*

As discussed, in-service inspection (ISI), surveillance programmes and maintenance requirements provide the template needed to maintain the original or sufficient margins in fission nuclear plant and components and to return them safely to service following plant outages. Fission programmes have featured post-commissioning surveillance to build up the confidence in the design, and to help establish the (mechanistically) based dose damage relationships (DDRs) for a wide range of material and irradiation variables.

Although the more rigorous regulatory environment will probably not allow un-supported use of DDRs for Safety critical equipment, they may be justified for equipment not fundamental to basic safety but nevertheless crucial to the R&D mission success of the fusion reactor. This is discussed in more detail in section 7. MAG therefore recommends that it is essential to develop at an early stage the remote ISI, material surveillance, maintenance and repair procedures to be used in-vessel on DEMO and to factor these into the design concept.

4. **Assumed Materials Loading requirements in DEMO**

   **4.1 Predicted neutron damage levels in DEMO**

   The levels of neutron flux and material damage in DEMO and a fusion reactor have been simulated in many papers. For this assessment, we took recent simulations from Gilbert et al [14, 15] using the most up-to-date cross-sections for the evaluation of damage levels and helium/hydrogen transmutation production. These simulations involved a model $2.7GW_{th}$ fusion reactor, similar in size to the Roadmap DEMO with a helium-cooled 'Pebble Bed' (lithium-based ceramic and beryllium multiplier) BB and a tungsten-surfaced divertor. The simulation shows poloidal variation of +/- 20% in the neutron flux through the first wall surface, with a peak of ~8.25 $10^{18}$ n .m$^{-2}$.s$^{-1}$, corresponding to a damage level of ~ 18 displacements per atom per full power year of operation (dpa/fpy) in a first wall steel. This was scaled down to the DEMO concept of section 2.1 (1.8GW$_{th}$) giving a damage level of ~ 12 dpa/fpy in the first wall steel of the DEMO BB. The simulations [14] also show a production ratio for transmutation helium of ~12 appm/dpa.

The increased focus on a more exact simulation of the neutron loading variation with breeding blanket concept and the need to re-assess the baseline concepts in the Roadmap has resulted in further simulations where the fluxes and damage levels have been compared for different BB concepts in the same machine [16]. This shows that neutron flux levels are, in general, higher with liquid metal (lithium-lead) BB concepts than in equivalent ceramic breeder concepts, due to the increased neutron moderation in the latter. The peak neutron flux in the simulations in [16] is ~10% higher than the figures in the previous paragraph (occurs for the helium-cooled lithium lead concept). This validates the conservative assumption of taking 15 dpa/fpy (and 220 appm/fpy helium concentration) as the damage level at the front face steel in the MAG assessment. The neutron irradiation load will vary strongly through the BB as the high energy flux is degraded through interactions causing damage and transmutation, breeding and multiplication reactions, and the simulations show [15] that the neutron flux will have dropped by an order of magnitude after traversing the first 0.25m of a composite helium-cooled blanket, and even larger drop would occur for water-cooled concepts. As a result the radiation damage will drop to the equivalent of a few dpa, and the risks associated with any particular structural material will change through the blanket structure (the rear of the divertor will be similarly shielded).

The neutronics simulations [15] scaled to the DEMO concept power level show that the neutron damage level in a PFC material like tungsten (as used in ITER) will be $\leq$ 9 dpa/fpy for first wall armour, and $\leq$6-7 dpa/fpy for the divertor, whilst for tungsten in the shielded plasma 'strike-zones' of the divertor, the requirement may be as low as 2-3 dpa/ fpy. The helium transmutation is much lower than in steel, with predicted levels $\leq$4 appm/fpy in the first wall, and $\leq$2 appm/fpy in the divertor armour. As with the steel values, we have taken a 25% margin in the assessment. This would be particularly important in assessing tungsten, as the recent (TENDL 2011) cross sections used in the simulations [17, 18] have led to a ~ three-fold increase in the tungsten damage predictions.

### 4.2 Loads on in-vessel structures

*4.2.1 Mechanical Loads*

The mechanical loads on in-vessel components in the Roadmap DEMO concept will have a cyclic component typical of a pulsed device. For the 20dpa lifetime of the concept DEMO's BB, a level of 15dpa/fpy corresponds to $\geq$ 1.33 full power years of operation in Phase I, and, with a 2-2.5 hour pulse time [4], the number of fatigue cycles will be $\geq$6 $10^3$, rising to $\geq$15 $10^3$ in the Phase II concept.. Structural material will require lifetime against creep-rupture of $\geq$12. $10^3$ hours at maximum stress and typical operating temperature for Phase I ($\geq$30. $10^3$ hours for Phase II). Maximum stress and operating temperature are, of course, intimately determined by the design. More will be discussed on these issues in section 5.

*4.2.2 Thermal loading and plasma erosion*

The loading for DEMO Plasma Facing (PF) materials is severe, especially in the divertor. Analysis of DEMO options reveals a complex situation [4], with power loadings as high as 50 MW.m$^{-2}$ at the plasma strike zones in the absence of plasma detachment, which acts to shield the power from focussed interaction with the target, dissipating the majority of the power via radiation. The highest fraction of plasma power radiated, achieved by impurity seeding, is ~90% [19], but maintenance of high energy confinement performance in the bulk plasma is an issue at these levels. Erosion by plasma ions is severe in the divertor region, where the fluxes rise by more than two orders of magnitude to $> 10^{23}$ m$^{-2}$.s$^{-1}$ [20]. With the impurity seeded plasmas, moreover, the erosion is dominated by introduced impurities, as discussed in section 6. We have taken a power loading of ≤20 MW.m$^{-2}$ as a guideline for the divertor strike zones. This requires some thickness of armour to avoid 'punch-through' erosion of the PF surface, and hence for the divertor, the PF armour must also have a high thermal conductivity, essentially acting as a HHF material too. The loadings for the first wall armour are presently assumed to be much lower, ~ 1 MW.m$^{-2}$ - 5MW.m$^{-2}$ on the basis of ITER estimates [21]. This is an area where further plasma configuration simulation of a DEMO concept is required [4].

5. **Structural Materials**

   **5.1 Baseline**

   The MAG assessment confirms Reduced Activation Ferritic Martensitic (RAFM) steels the Baseline Material choice for the BB structure. Within the EU programme, the reference material is EUROFER [22].This Fusion programme development has the following strong-points:

    - good overall balance of mechanical properties required (high strength, ductility, fracture toughness, high fatigue cycles at relevant strain levels);

    -sufficient corrosion resistance to liquid (LiPb) metals in breeder concepts at the relevant low flow conditions in the reference EU blanket (helium-cooled lithium lead,

where the liquid metal is used as breeder rather than coolant) in a useful temperature range (up to ~550°C) [23, 24, 25]; and

- sufficient compatibility with He-gas cooling, making it compatible with the He-cooled EU reference blanket concepts [4].

EUROFER's development is in the TRL4-5 range, and significant quantities have been fabricated industrially, welding developed and fission-irradiated properties measured quite extensively, and in addition, EUROFER, as the structural material chosen for the ITER Test Blanket Module (TBM) programme is being integrated into the RCC-MRx nuclear code. Crucially, it has the relatively good stability under fission neutron irradiation of a body-centred-cubic (bcc) latticed material, with very low swelling.

*5.1.1 EUROFER Risk Log and mitigations*

The Risk Log for EUROFER shows many issues however many of these should be dealt with in the ongoing development programme. Below, we concentrate on the most serious, with 'very high' or 'high' impact on the DEMO design, systems engineering/ plant operations or project mission success.

*Highest impact engineering design risks.* EUROFER's properties, and their development under irradiation, lead to a limited temperature operating window, restricted to particular BB designs with FW temperature ranges between 350 and 550°C [22]]. The most serious risk comes from low-temperature embrittlement under fission neutron irradiation. The exact temperature limits are uncertain because of the un-quantified added effect of the fusion-neutron produced helium embrittlement, but safe operation of the blanket should only be guaranteed if irradiated above 325 - 350°C operating temperature, as this avoids a shift in the Ductile-Brittle Transition Temperature (DBTT) levels (ie. DBTT ≈ 0°C or above) which would be unusable in an operating reactor [26, 27]. There is evidence that EUROFER's irradiation damage can be annealed-out by post-irradiation high temperature (550°C) cycling [28], but this operational cycle could be difficult to achieve in a realistic design, and any 'ratchetting' effect of DBTT-shift after many cycles is unknown. This mitigation is worthy of more detailed study.

Another serious risk in this category concerns the lack of validated understanding of effects of transmutation helium on embrittlement and swelling due to the absence of an irradiation facility with a fusion-relevant neutron spectrum. The effect of the mobility of transmuted helium on the possible annealing cycles mentioned above is also an unknown. Surrogate simulation techniques are currently used, such as fission reactor irradiation of $^{10}$B-doped steels, or He ion-beam implantation data, and these may provide useful estimates of helium embrittlement effects, but many fundamental issues are involved, surrounding such surrogate experiments, some of which are discussed in section 7. As an example of the possible effects, for He ion-beam implantation, data exists [29] showing that at ~30 dpa the properties begin to exhibit radiation hardening as He concentration exceeds ~ 500 appm. This limit is only ~ 50% above the specified limit for DEMO Phase I, and the He concentration is only double the specification. These narrow margins indicate the need for a co-ordinated campaign to obtain better data and understanding, to mitigate the risks.

Finally, the lack of Design-code development (both Structural Design Criteria and Regulatory Codes) is currently a high-impact risk on engineering design. Some mitigation of this will come from the developments of TBM codes for the RAFM structures of the ITER Test Blanket Modules [30]. More mitigation measures are needed, associated with the treatment of the DEMO Safety Case and fusion neutron spectrum testing (see section 7).

*Highest impact Systems engineering/plant operations risks*. EUROFER's strength declines above the 500-550°C range, where the creep-rupture failure also drops below $10^4$ hours, for stress levels of ~ 100MPa similar to those proposed as maximum primary stress levels for fusion reactor developments of fission codes such as RCC-MRx [31]. It is to be noted in this respect that the allowed operation temperature limit in such codes as ASME for the creep-rupture lifetime for turbine steels is usually quoted as that at which a lifetime of $10^5$ hours at stresses of order 100MPa is achieved [32].

These limits clearly entail design risks for DEMO, but, equally important, the attendant risks involve systems engineering and plant operation risks for the reactor design. The low-temperature embrittlement, coupled with the decline in strength gives a difficult, relatively narrow, temperature operating window for a Breeding Blanket and make it difficult to envisage a high-temperature coolant loop (with high thermal efficiency) with a EUROFER blanket as its 'front end'. This runs counter to the use of helium cooling, as the coolant gas cannot be used at a temperature range where the Balance of Plant can employ the potentially higher thermodynamic efficiency of a Brayton cycle, or even reach the upper limits of a supercritical Rankine cycle, both of which require operation temperature in the region of 650°C or greater [33].. Studies of BoP using the helium coolant temperature limits arising from EUROFER indicate [34] that the thermodynamic efficiency of Rankine or Brayton cycles limited to < 480-500°C are insufficient to reach the net electricity generation target for the DEMO concept when the high installed pumping power for the helium circulation (~150MW) is subtracted. This entails a high-impact risk to the DEMO mission. Another, potentially high-impact risk arises from the limitation that the EUROFER upper operation temperature places on any annealing cycles used to restore

EUROFER ductility post-irradiation, as discussed above. This factor needs more study in the R&D to investigate this mitigation measure.

*Highest impact DEMO project-level risks.* The overall risks to the DEMO mission from the use of helium-cooling technology, where circulators and compact heat exchangers for high-pressure helium are immature technologies [34], motivate a strategy of parallel blanket concept development to accompany the baseline ITER TBM helium-cooled programme [4]. Water-cooled systems have the attraction of mature, low-risk BoP options essentially similar to those of fission Pressurised Water Reactors (PWR), and BoP studies [34] show that even simple systems come close to the required overall plant efficiency. EUROFER's embrittlement temperature makes its use problematic in a water-cooled blanket such as the Water-Cooled Lithium Lead concept, with 290-320°C operating temperature. Indications exist that some melts of RAFM steels have had superior embrittlement properties around 300°C, especially the EUROFER 97WB [35], and the Japanese F82Hmod3 [36]. For risk mitigation the MAG assessment recommends a development programme to push RAFM to lower embrittlement temperature. [1]

As with all 8-9% Cr FM steels, corrosion under irradiation and stress-assisted corrosion cracking would potentially be high impact issues if water were to be used in a BB [39], and coating and coolant chemistry mitigations will be required. A thorough review of the extensive lessons learnt from the fission reactor programme on this topic [40] is recommended, with assessment of their application to fusion.

---

[1] The embrittlement temperatures quoted are the standard ones derived via high strain rate (Charpy) impact tests. A detailed discussion is beyond the scope of this paper, but we note that the tendency for brittle fracture to occur depends not only on temperature and the inherent toughness of the material, but also on other factors such as strain rate and degree of constraint (a factor particularly important in the case of thin sections). A Master Curve approach [37] can be used to provide a conservative estimate of the likelihood of failure for blanket designs under particular operating conditions. In addition, there have been developments in fusion-specific codes, primarily the ITER SDC-IC [38] to define limitations to extrinsic factors such as secondary stresses (as induced by eg. thermal cycling) and peak stresses (as induced by, eg. design shapes and support constraints), in order to mitigate against the effects of loss of ductility. See also the discussion in section 7.

References such as [40] emphasise the importance of probabilistic fracture analysis, fracture-mechanics based studies and in-service inspections.

**5.2 Blanket Structure risk mitigation – high temperature steels**

The central importance of the blanket structure, makes an active risk mitigation programme for the structural steels necessary, and the likelihood of high-impact risks remaining post mitigation makes a programme on Risk Mitigation Materials essential, either as complete replacements for EUROFER, or to complement the use of EUROFER in zones of high irradiation. Two existing lines for RMM are recommended:

- developing fusion-compatible high-temperature Ferritic-Martensitic (HT FM) steels from those developed outside the fusion programme with improved high temperature creep strength (up to ~ 600 - 650°C); and

- pursuing the development of oxide dispersion strengthened (ODS) alloys, subject of fusion R&D for a decade, but still without industrialisation. These have good high temperature tensile strength and creep resistance.

Several other alternatives to RAFM steels as structural material for the blanket have been put forward. Amongst the 'advanced' materials, vanadium alloys and SiC/SiC composites have been most frequently mentioned. Our assessment does not recommend the high priority pursuance of these options, although they may be of interest as 'advanced' second or third generation materials. Vanadium alloys (principally V-4Cr-4Ti) have been investigated for fusion, but their low-temperature embrittlement limit is higher than EUROFER (~400°C), whilst their creep-rupture limitations lie in the ~650-700°C region [41, 42, 43], thus their temperature operating window does not open up significant advantages for design flexibility. In addition, they suffer from embrittlement when in contact with hydrogen and impurities (such as

C and O),[41, 44] and have very high solubility for tritium (a problem both for the plasma facing and the breeder facing surfaces of a vanadium blanket concept) [45].

Whilst SiC/SiC composites have been widely quoted as possible structural materials in reactor studies such as the US ARIES series [46], there remain significant problems with SiC/SiC in this context. The material is only 'quasi-ductile' (crack tolerant) at best, and has a limited bending strength. Importantly for a heat-removal system, such as the blanket, its thermal conductivity deteriorates rapidly under irradiation (for the high thermal conductivity variants this occurs by a factor ~ 30 if irradiated at 200°C up to ~ 0.1dpa, and still by an order of magnitude if irradiated at 800°C for levels up to 2-3dpa[47, 48]). SiC/SiC is also prone to swelling (>1% volume at 1 dpa for temperatures below 600°C), and whilst this level of swelling would not rule it out as a structural material, it is expected to be enhanced by the very significant levels of helium transmutation simulated to occur under a fusion neutron spectrum (>10000 appm is expected at the blanket first wall over a 5 year lifetime [49, 50] ).

The use of solution annealed 316L-(N) stainless steels as RMM for BB structural steel was rejected. Although previously studied as a potential candidate blanket structural material for fusion reactors, its power exhaust capability is below that of RAFM steels and might be an issue for a DEMO BB. Crucially, these steels' resistance to radiation effects is limited. Their ductility and fracture toughness are severely degraded during irradiation around 300°C and damage of ~10dpa [51, 52], whilst rapid failure occurs at around 330°C at only 3-4% elongation after 7dpa irradiation [53]. At mid-range temperatures of 400-500°C they are susceptible to unacceptable volumetric void swelling for doses >20 dpa, and they have been shown to suffer from severe He-embrittlement at high temperature (>550°C), in slow strain rate testing, for He contents above 10-100 appm (well below the ~ 220 appm DEMO 'Starter Blanket' FW conditions).

The chosen Structural RMM types have reached reasonable levels of development (TRL3-5, the high temperature FM steels largely in advance of this if un-irradiated behaviour is considered), and their advantages and drawback issues can be identified, as discussed below. Figure 2 shows how the US developments for both lines have demonstrated much higher tensile strength than conventional RAFM variants.

*5.2.1 ODS Steels*

The ODS programmes intend to *complement* RAFM steels with a structural material aiming at both higher temperature and improved irradiation resistance, and with dispersed oxides intended to provide precipitate sites to 'fix' the helium gas bubbles generated by high-energy fusion neutrons thus preventing movement to grain boundaries and enhanced embrittlement. Although tubes and sheets have been successfully produced, ODS steels currently have the following drawbacks:

 - the experimental batches produced typically have low fracture toughness at room temperature, implying further basic development, although some improvements have been made with 'second generation' heats with extended high temperature, post production heat cycles [55] ;

 - fabrication of components will be difficult as the current materials have anisotropic mechanical properties unless a complex thermo-mechanical treatment (TMT) is followed; and

 - the quality of the experimental heats is highly variable.

Moreover the steels are only available in small (kg size), laboratory made quantities, and the process for manufacture, with much powder metallurgy and intensive, complex TMT, will have to be scaled-up and industrialised. It is worth noting that there have been, as yet unsuccessful attempts to manufacture ODS in bulk without resort to powder metallurgy. Other serious issues for ODS are the lack of welding techniques available (only friction stir-welding looks promising), and the poor knowledge of the non-irradiated engineering parameter database in the case of 12-14Cr versions of these steels.

The ODS steels also share the low-temperature radiation embrittlement problem with EUROFER. There are some experimental indications however, that 9Cr ODS steels exhibit less radiation-induced hardening asymptotically than conventional RAFM steels [56], and that hardening and DBTT shift are negligible up to ~ 1.5dpa in 14Cr ODS [57]. These effects are attributed to the higher density of precipitate nano-clusters, improving the dislocation density and providing sinks for radiation defects, thus less severe low-temperature radiation embrittlement is expected.

For ODS it is of crucial importance to engage with industry to discuss the industrialisation of production processes, and establish robust welding techniques. This effort needs to be pursued in parallel with the more fundamental experimental and modelling research, as there are considerable hurdles to overcome.

*5.2.2. High temperature Ferritic-Martensitic steels*

The current 'industrial standard' FM steel in the power generation field is Type 92 (9Cr, 0.5Mo, 2W) [32], a development of the Type 91 from which the EUROFER-type RAFM steels were developed. This has a creep-rupture performance of $10^5$ hours at 100MPa and 620°C, but this drops below $10^4$ hours at 650°C. This steel type is clearly at TRL8-9 for non-fusion applications, with full inclusion in codes and design standards, and even development of this for reduced activation would yield benefits. Indeed RAFM steels developed by some of the 'newer' members of the worldwide fusion research community have moved the basic EUROFER/F82H type of RAFM steels towards a 2% tungsten content, with some demonstrated improvement in creep and tensile strength properties [58].

The goal for DEMO Phase 1 BB is >1.2 $10^4$ hours at temperature, and for a reactor the EU PPCS fusion reactor studies [59] identified a 5 year blanket lifetime as an economic goal (~5 $10^4$ hours). Thus it can be seen that the industrial standard HT FM steels do not meet these targets. However, the extensive world-wide R&D into improved HT FM steels is producing a 'fourth generation' of FM steels showing excellent promise at high temperature. The driver for this comes from the developments required for ultra-supercritical steam generator fossil-fuel plants. Two approaches were considered by MAG: the steels subjected to special rolling treatments during the plate-production process (thermo-mechanically treated – TMT – steels); and those subject to an alternative heat treatment cycle (AHT steels). In the TMT steels process, the 25mm thick plates were subjected to a series of experimental hot rolling trials. The precipitates induced by rolling, formed along the dislocations, and results were achieved (figure 3) with a 1000-fold increase in precipitates (0.2-1. $10^{22}$ m$^{-3}$ at diameters of 4-8 nm), refined prior austenite grain size and much increased density of sub-grains [60, 61]. The increased tensile strength which results is shown in figure 2. Amongst this new HT generation of TMT steels, the creep-rupture performance still needs proving beyond ~ 3 $10^3$ hours at 650°C[62], however *AHT type 92 variants* have reached > 3 $10^4$ hours at 650°C and 92MPa[63], indicating the promise of these steels.

The AHT FM steels have been developed with a higher austenitising temperature (up to 1200°C) and a lower tempering temperature (around 660°C) than in the usual manufacturing process for type 92. The aim being to both maximise the alloy in solution and increase the precipitation nucleation rate, thus maximising precipitate

density. The melts produced have the following features[64]:

- reduced $M_{23}C_6$ levels;

- finer Z-phase (Cr{V, Nb}N) precipitate;

- higher density of MX precipitates and an improved balance of MX:Z-phase; and

- increased density of $M_2X$ precipitates.

Finally, the authors of [64] have increased the nitrogen:carbon ratio in the steelmaking process approximately doubling it above the typical type 92 values . This has yielded the best creep performance, as shown in figure 4. The latest results for 650°C, show that the nitride-rich AHT steels have very nearly reached the 5 years (~4.4 $10^4$ hrs) without failure. The use of nitrogen to strengthen steels is well known in the wider industry and is also seen in fusion steel development, eg. in the Japanese programme. The creep-rupture performance of F82H RAFM steel is seen to improve by approximately an order of magnitude in tests at 550°C [36]. There is a potential issue with nitrogen addition, as, studies [36] show that the activation from the BB waste will be dominated by nitrogen activation products (principally $^{14}C$) after 100 years, though not necessarily to the extent that the low-level waste classification so important to RAFM steel justification will be endangered.

With the current state of R&D, the TMT- or AHT-modified FM steels have the serious issues of:

- very limited development of reduced activation variants for the HT FM steels [54, 66], and no development at all in the case of the AHT FM line;

- lack of fission irradiation data and data on helium transmutation embrittlement;

There is a potentially serious issue that there will be unacceptably-reduced creep properties following welding for these steels, as the relationships between the post-weld-heat treatment (PWHT) temperatures and the tempering temperatures can be unconventional. However, there is evidence (eg. ref [64], table 6) that this is no worse than in conventional ferritic-martensitic steels, and PWHT joints can be realised with >75% of the parent material creep strength..

These steels are expected to exhibit low-temperature embrittlement problems, but, on the positive side, their high density of nano-scale precipitates and microstructures are predicted, on physical metallurgy grounds, to lead to superior performance to EUROFER on low temperature and helium embrittlement [67]. This prediction also holds for ODS steels [eg. 68]. As an example, figure 5 shows the onset of radiation hardening is delayed in existing RAFM steels, with their relatively high dislocation sink-density, over the relatively low sink-density bainitic reactor pressure vessel (RPV) steels [69]. There is additionally an improvement in the rate at which the DBTT change occurs in the RAFM steels as a function of the irradiation hardening [70]. These coupled facts give a strong motivation in the direction of inclusion of these steels in the R&D risk mitigation programme. There are initial indications that the nano-precipitates in the ODS steels will re-form after neutron damage [71], and this positive sign needs further exploration for the other steels. The modelling of this process at the fundamental damage level, coupled with the more macroscopic considerations from the Computational Thermodynamics of the solubility of precipitate products in the steel matrix (which will determine the propensity of nano-scale precipitate fragments to dissolve rather than recombine) would be an important input to this work.

As 'classical' steels the industrial fabrication and welding development of TMT and AHT high temperature FM steels should be relatively straightforward, and has already reached a relatively mature stage The MAG recommends that the development of Reduced Activation HT FM steels be pursued in close co-operation with industry, as common goals exist in the development and industrialisation of these steels. The highest priorities for this development are the production of RA versions of the TMT and AHT steels, and obtaining data on the irradiated properties of the existing varieties.

## 6. HHF/PFC Materials

### 6.1 Selection of water-cooled divertor as baseline

The choice of concepts for divertors lies between low-temperature (100-200°C coolant) water-cooled concepts, of the generic type used for ITER [72], and high-temperature (600°C+ coolant) helium-cooled concepts [73]. The latter are at a much lower level of development (< TRL3), and limited to power handling < 10 MW.m$^{-2}$, whereas the former are at ~TRL4, and tested to ~ 20MW.m$^{-2}$. In general water-cooled systems can reach much higher heat transfer values than gas-cooled systems, and hence the near-term concept DEMO selects a water-cooled Divertor for the Baseline [1, 4]. The safety issues of having water cooling in the Divertor, combined with a liquid Pb-Li breeder in the blanket are minimal in nature. Both systems, from the viewpoint of mitigating an individual in-vessel Loss of Coolant (LOCA) event, will have to be engineered so that the water and the liquid Pb-Li are doubly-contained in their respective systems. This results in a four-fold barrier between the two coolants in-vessel. A four-fold simultaneous rupture of confinement is an event beyond the Design Basis Accident (DBA), and consequently would not be a major issue either in formal Safety Case approval, or in practical operation.

### 6.2 Baseline PFC and HHF materials

Tungsten is regarded as the Baseline Material for state-of-the-art plasma-facing component technology, and this is confirmed by the MAG assessment. The key advantages of tungsten are: the high threshold energy for sputtering, around 100-200 eV by hydrogen isotopes; and the low retention of tritium in the material, important both for the feasibility of the D-T fuel cycle and the Safety Case for fusion reactors, which, on ITER precedent, is keenly concerned with the inventory of mobilisable tritium in case of any loss of vacuum accident (LOVA).

Copper alloys are recommended as the material for the HHF heat sinks in the water-cooled divertor design, due to the unrivalled heat conduction of copper, and the existing tungsten-copper alloy design for ITER, which aims to prove a working design tolerant of the brittle nature of tungsten. Copper alloys, and especially Cu-Cr-Zr, are well documented and understood as structural materials.

*6.2.1 Tungsten and copper alloy risk logs and mitigations*

Using the previous categorization, the highest risks relating to tungsten as a PFC material are mainly in the categories of systems engineering, operational and even DEMO project-level and hence demand serious mitigation programmes. There are high-impact design engineering risks in use of both tungsten and copper and theseimpact on the design of the divertor as an *engineering system* in itself, especially applicable to a low temperature water-cooled design.

*High impact design engineering risks in the use of tungsten and copper alloys*

Tungsten is an intrinsically brittle metal, with a DBTT ~ 500°C, even in an un-irradiated state. Moreover there is a lack of radiation embrittlement data for tungsten, and the combination of these facts poses high design engineering risks. The un-irradiated materials database for tungsten also has some serious shortcomings (eg. thermal fatigue data) if tungsten were used structurally, as opposed to PFC armour. High priority R&D programmes must resolve these last two shortcomings and engineering work-arounds are required to address the first.

The most serious issues for copper alloys relate to the rapid loss of ductility under irradiation at temperatures < 180°C, and the alloys' operating temperature for DEMO should thus be kept above 200°C. In addition for some alloys, i.e. Cu-Cr-Zr, a combination of normal over-ageing [74] and the decrease in strength with irradiation, which sets in above 250°C, limits the upper temperature for engineering structural applications (eg. in the ITER SDC-IC code for in-vessel components [38, 75) to 350°C for doses up to ~5 dpa. Design use of copper alloys (without better composites) may have to be restricted to substructure and coolant systems in the *immediate Divertor strike zone vicinity,* where neutronics simulations [14] show that damage is reduced by a factor ~ 3 compared to the outer edges of the divertor (figures ~ 3 dpa/fpy are predicted, as shown in figure 6). The strike zones and surrounds are where the high quality of the thermal conduction and material strength must be maintained, at the outer regions of the divertor, the heat fluxes are more typical of the rest of the first wall poloidal cross section.

*High impact systems engineering and operational risks for tungsten*

The highest impact risks in this category come from:

- uncertainty in the erosion effects on neutron irradiated tungsten materials (effects of neutron-induced defects).;

- potential oxidation/deflagration of tungsten PFC, either as a blanket covering or divertor surface in an up-to-air accident scenario – this is a licencing issue for DEMO, and arises because, unlike in ITER, there will be significant decay heat present in the transmutation products of the activated first wall structure;

- damage by plasma edge instabilities ('ELMs'), where energies ~ $GW.m^{-2}$ impinge on the divertor in ~ms periods.

The first risk includes the production of 'tungsten fuzz' during bombardment by helium ions [76], which form ~ 10% of the content of afusion reactor plasma. To quantify the effects of irradiation there is an urgent need for plasma stream experiments on neutron-irradiated tungsten.

To mitigate the deflagration risk, there is an urgent need to develop self-passivating tungsten alloys. Considerable progress has been made in this research already [77], with factors ~ 1000 suppression of the oxidation rate achieved for alloys such as WSi6Cr10Zr4. It will be necessary to model the transmutation of such alloys under neutron bombardment, and to establish their tritium retention characteristics before final decisions are made.

The damage threshold for tungsten under an intense short bombardment such as encountered in an ELM-event is ~0.2 GW.m$^{-2}$ [78], and one of the major risk mitigation measures identified by MAG, and the subject of world-wide research effort, is the development of high-performance plasma regimes where ELMs are minimized or eliminated [79, 80]. Many of the plasma regimes feature significant impurity injection to limit the power flux to the divertor and to control ELMs, and these are now combined in machines such as ASDEX-U and JET [81, 82] with the need to control tungsten influx into the plasma. These impurities impinging on the divertor lead to greatly enhanced sputtering rate (as shown in figure 7) and the requirement to keep the plasma temperature to ~ 5 eV in front of the divertor, a very difficult target [81]. The coupled nature of the problem, and the interaction of the plasma and divertor system as a whole can be appreciated by noting that the 5eV limit corresponds to only 5 MW.m$^{-2}$ power loading, thus running counter to one of the main arguments for adopting the water-cooled divertor. Consideration of figure 7 shows that this limit could be lifted if nitrogen seeding were not used. As the improvement in performance of the 'tungsten-walled' plasmas which comes from nitrogen seeding does not yet extrapolate without caveats from ASDEX-Upgrade to JET, the tokamak

R&D is far from finished for extrapolation to DEMO. Thus one major risk mitigation measure for DEMO, identified in the Roadmap is to develop plasma regimes consistently integrated with divertor technology for DEMO. This will require not only experiments in ITER, but in other machines, possibly with a dedicated 'Divertor Test Tokamak' being required [1]. Possible solutions exist in novel magnetic configurations for the divertor, to assist in the power loading spreading, and hence erosion spreading, of the strike zone area, but these are not yet demonstrated as compatible with high-level plasma confinement performance. An ongoing iterative process of establishing tokamak regimes and then examining for material risk levels is foreseen in the Roadmap R&D.

*Highest impact DEMO project-level risks for tungsten*

The most serious project-level risk involving the use of tungsten, or its alloys, is the lack of any data on the irradiation effects on the tritium retention by tungsten. There is a clear safety and licencing issue in this respect regarding the minimisation of mobilisable in-vessel tritium inventories in the event of a Loss of Vacuum Accident (LOVA). Such issues, established by the precedent of ITER, will be firmly in the mind of regulators licencing future DEMOs [11]. A further consequence of enhanced tritium retention in tungsten however, would be the degradation effect on the tritium breeding cycle of tritium hold-up in the vessel surfaces, and the potential for outages whilst the in-vessel surfaces were too frequently de-tritiated. These three consequences of potential irradiation-enhanced retention constitute a threat to the DEMO mission itself. As such a major R&D effort is required, including a plasma-stream facility capable of handling activated, irradiated tungsten samples, as indicated above for the erosion measurements. Deployments of experimental samples will, in this case, have to be preceded by extensive modelling to investigate the possible interaction between the implanted tritium ions from the plasma, and the neutron damage. In particular it is very relevant to model accurately the temperature behaviour of the system, as the affected layers of the tungsten will, even in a water-cooled divertor, be operating at high temperatures during implantation.

*6.2.2 Illustration of operational and design problems by simple systems engineering of the divertor concept.*

To illustrate the systems engineering design challenges of a low-temperature tungsten-CuCrZr water-cooled divertor, we show a schematic cross-section in figure 8. For a power handling target of 20 MW.m$^{-2}$ and the average thermal conductivities for the materials over the likely temperature operating range (note for tungsten these are un-irradiated values), the temperature limits to avoid certain performance-degrading effects, based on current understanding, are shown. One potential surface temperature limit for tungsten comes from avoidance of re-crystallisation (this occurs above ~ 1200°C in pure tungsten, and is complete after ~ one hour at 1300°C). Re-crystallisation is measured to cause a significant reduction in toughness. The mitigation for this risk comes either in the development of various alloys of tungsten, all of which are seen to raise re-crystallisation temperature [83], or, if feasible, thermal stress relieving designs such as castellations etc. There is an extensive research programme ongoing on tungsten alloys [84], and mechanically promising candidates will need further qualification for properties such as hydrogen-isotope retention, plasma erosion and post-irradiation activation products. A potentially more serious limit comes from the minimisation of 'surface fuzz' (which becomes significant, in comparison to the 'normal' plasma erosion above ~900°C [76]). The experimental evidence is gained to date in exposure to helium plasmas, and to quantify this further, in realistic plasma exposure (hydrogen ions with a ~ 10% helium content) needs plasma stream experiments. It is important to see if there is any synergy between the fuzz growth and the 'normal erosion', or conversely any beneficial impediments. Identification of any potential synergy is important, as its existence may lead to an unfeasibly thin tungsten layer (see below). To avoid the most serious irradiation-induced swelling (~ 1-1.6%) observed at ~ 10dpa levels [85], the bulk temperature of the tungsten should be below ~650°C; whilst to operate the tungsten in a ductile regime, the bulk should be above ~ 600°C. For the copper alloy meanwhile, the previously discussed limits for operation lie between 180 -350°C. The copper alloy limits and the tungsten surface limits translate, using the thermal conductivity values, and the assumption of water-cooling at ~ 150°C, as in ITER, to thickness limits of ≤2.8 mm for CuCrZr and ≤3.5-4.8mm for tungsten (the lower limit coming from the 'bare' results on tungsten fuzz production). These represent feasible engineering structures to last for 2 fpy in operation unless erosion is significantly enhanced by radiation. However, the *bulk* temperature limits for tungsten cannot be satisfied, showing that, unless new materials are forthcoming, design engineers will have to cope with brittle materials and significant swelling. In passing, we note that the restriction of the divertor coolant to these low temperatures still enables the divertor outlet water to be used to pre-heat a Rankine cycle for a water-cooled breeding blanket and thereby gain in overall DEMO plant efficiency [34].

### 6.3 Advanced HHF/PFC Materials

Even with substantial progress in the R&D programmes proposed, there are likely be remaining residual high-impact risks. The clear 'show-stopper' nature of the divertor development makes it essential to develop an RMM programme for HHF and PFC Armour. Many promising candidates have been identified for possible divertor/first wall applications:

- fibre and foil reinforced composites of copper and tungsten [86, 87];

- tungsten laminates [88];

- tungsten-copper functionally-graded materials.

Notwithstanding promising results [77], these are at a very early stage of development (TRL 1-2), and there is a complete lack of data on irradiation results. The MAG proposes a generic materials R&D programme to develop towards a down-selection before the end of Horizon 2020. We also note the importance of tritium permeation barrier development, associated with an integration of the barrier layers in a first wall component, and that ceramic barrier layers must be developed as an integral part of a first wall component.

7. **Materials irradiation testing issues**

As they will suffer from helium-transmutation production, it is clear that data from irradiation under a 'Fusion neutron spectrum' is essential as a precursor to the final engineering decision on DEMO materials. In order to plan this correctly, and ensure that the timely execution of the project is not endangered, it is essential to analyse in more detail what the true 'prior- licencing' aspects will be for DEMO materials, ie what will be required of regulatory Codes and Standards, and what the role of design engineering requirements such as Structural Design Criteria (sometimes confused with licencing requirements) will be. Splitting the materials data problem into these areas, as is recommended by examination of the lessons from fission, and the ITER Safety Case (discussed in section 3) will enable a solution to the problem (previously faced in the fission programme) of 'bootstrapping' the concept DEMO into a position where regulators and funding stakeholders, and the wider community, are satisfied to go ahead. If the relatively-lightly irradiated Vacuum Vessel is chosen as the primary boundary, with passive safety and engineered 'defence in depth' provisions, its irradiation embrittlement should be low, with damage levels < 1-2 dpa, even after 30 full-power years' service [33]. The very much softer neutron spectrum [14, 15], with negligible fluence above 1 MeV will produce very much reduced levels of H and Fe transmutation, and the He-content is predicted to be limited to ~$10^{-4}$ appm/fpy, thus normal fission-spectrum testing is sufficient, and could go ahead at an early date. The MAG recommends that ITER practice in this respect is carried over to DEMO, with Safety Analysis to establish the primary containment boundary for DEMO at an early stage in design. With such low fluence, the DEMO VV could even be constructed of similar material to ITER (Austenitic 316L) and, with careful shielding design, the radioactive waste at end-of-life could still be lower than that from a first wall blanket steel constructed of Reduced Activation steels developed for fusion. The potential for this is shown (figure 9) in a simulation of the time-histories of the specific, or total, activation and dose-rate of the outboard mid-plane vacuum vessel constructed of 316L 'ITER grade' stainless steel after a 30 fpy lifetime (assumed end of life), compared to the EUROFER first wall of a helium-cooled pebble bed (HCPB) blanket design, after its target in-service lifetime of 5 fpy exposure. The reactor and neutronics model is that used in references [14-16]. The results show that, 100 years after the exposure has ended, for this HCPB concept (one of the two baseline EU-concepts for DEMO), the dose rate from the stainless steel is nearly identical to that from the EUROFER, while the activity is significantly lower. We note however, that the activity levels at 100 years are very sensitive to the original carbon and nitrogen content of the steels because the production of $^{14}$C dominates the activity of EUROFER, while 316L, with much lower C and N content in the current simulations, has activity dominated by $^{63}$Ni. On the other hand, $^{14}$C does not contribute to the dose rate, and in this case the main contributions to the predicted Sv.hr$^{-1}$ level at 100 years come from $^{60}$Co in EUROFER (with $^{53}$Mn dominating at longer times), and $^{94}$Nb (mainly produced from Mo) in the 316L. We note additionally that the detailed results depend on blanket concept [16], but for all concepts simulated (two helium-cooled and two water-cooled) the relative results show the same favourable comparison. The remainder of the in-vessel DEMO materials should then be handled in a manner similar to ITER via a DEMO version of the ITER SDC-IC [75]. Analysis of this split shows that it is a template for fruitful early progress in the 'qualification' of materials and the development of the sort of engineering codes required for stakeholder funding commitments [90]. This reduces the licencing burden of the '14MeV' neutron test data, but development of suitable engineering codes for design, and stakeholder-funding and regulatory requirements in general will still require the 14MeV tests. The Roadmap milestone for provision of this data to match the Early DEMO construction decision in the early 2030s is to achieve at least 30dpa damage in steel samples by 2026. At this level, as previously noted, the helium-transmutation effects are beginning to be measurable. The MAG recommends that the Codes and Standards milestone for Early DEMO should be 2028 in the Roadmap for DEMO, but the exercise for the relevant materials should be started in Horizon 2020. Fusion spectrum irradiations will dominantly be from small size samples, and any use of these within standards such as RCC-MRx and ASME needs endorsement and adoption as soon as possible.

The IFMIF project [91] has a central place in the programme to test fusion materials, but with the analysis above, the extent of this data needed before a DEMO construction can fall short of the full 'qualification' of materials. Long-term, it is clear that Fusion materials development must continue, and therefore the full-IFMIF realisation should be kept as an ultimate goal, but the MAG, and Roadmap note that the full-IFMIF development is too ambitious to be accelerated to meet the required milestones. The Group therefore believes that an early start on a more basic 14MeV accelerator should be made. It would be optimum if this basic accelerator could be developed into the full-IFMIF at an appropriate stage (ie. during the DEMO Phase 1 construction). Proposals exist for lower power, reduced irradiation volume (50-100 cm$^3$), accelerator-based 14 MeV sources which would provide the 30dpa data levels by 2025 [92, 93], and the recommendation is for a technical assessment of alternatives, including the evaluation of the irradiation volume for the test materials. The IFMIF high-flux irradiation volume (500 cm$^3$) is based on a complete evaluation of all relevant parameters, so the evaluation needs to prioritise amongst these, in a risk-based manner.

Analysing the options for optimising a 14MeV testing programme, by extensive pre-cursor R&D with fission neutrons, or high-energy ion implantation experiments, we note that fission neutrons provide a reasonable simulation of the displacement-type damage to be expected at fusion neutron energies as the fraction of *surviving* displacements after damage is approximately the same in the two cases [94], whilst this is not the case for high energy ion beams. In general, fission neutrons do not produce copious helium, the exceptions being for the reactions:

$^{10}B(n,\alpha)^{7}Li$; $^{54}Fe(n,\alpha)^{51}Cr$; and $^{58}Ni(n,\gamma)^{59}Ni(n,\alpha)^{56}Fe$

all of which have been exploited in experiments with isotopically- and chemically-tailored steels. For PFC/HHF materials, He-ion implantation experiments are required. There are many issues relating to interpretation of results from these experiments, and a recent detailed review deals with these [95].

The ion-implanted data generally is restricted to near-surface volumes, and interpretation and testing of these is part of the subset of the overall problem of relating small-sample testing to the normal materials test samples. The issues involving the tailored steels include, but are not limited to:

- excessive rate of helium production in the irradiations compared to the rate of displacement damage (known to give unrepresentative results on helium migration);

- the effect of transmutant lithium on embrittlement;

- the limitation of tailored steels to a few dpa upper limit;

- the known metallurgical effects of B or Ni on precipitates.

The technical resolution of these issues and the role of modeling must be evaluated, as the experimental tests in the literature are now rather old [references cited in 95]. In particular more modern theoretical treatments of displacement damage evolution and helium production and accumulation at grain boundaries, can be challenged by isotopically-tailored experiments (from fast reactors to tailor the ratio of helium and displacements rates to more fusion-like values) for ferritic-martensitic steels where the defect sink strengths are radically different (as is the case between EUROFER and the new TMT FM steels).

It is clear that modeling will play a fundamental role in the optimization of the programme. The MAG report contained more general recommendations on Materials Theory and Modelling, and the reader is referred to ref [33] for a discussion of these.

8. **Conclusions**

The EU Fusion Roadmap exercise, with its emphasis on a DEMO concept for which a construction decision would be made in the early-2030s allows a sharp focus to be applied to the issues of materials development. The adoption of the project-based methodology allows risk analysis to prioritise the R&D programme needed to produce materials to maximise the success of the DEMO mission. This is aided by the treatment of systems engineering issues, and analysis of materials systems, and applying lessons learnt in the fission programme, especially in the development of safety cases, codes and standards.

A set of *Baseline Materials* has been identified for blanket structural applications (RAFM steels), plasma-facing components (tungsten), and high-heat flux materials (tungsten and copper alloys). The risks attendant in using these can be addressed via identified R&D programmes, but to be more certain of an optimum portfolio of materials, the development of *Risk Mitigation Materials* has also been identified for initial parallel development: high temperature FM and ODS steels for the structure, and composite tungsten and copper materials (especially laminate, fibre-reinforced materials and functionally-graded materials) for the PF and HHF applications. Materials testing with a fusion neutron spectrum remains a high priority, but using the ITER precedent and fission experience, the vacuum-vessel of DEMO can be identified as the primary safety barrier, and this can be tested with a fission spectrum in existing machines, thus reducing the burden on fusion spectrum testing. The acceleration and optimisation of a fusion spectrum test programme is recommended via the early deployment of a less powerful 14 MeV neutron source compared to IFMIF, and by the pursuit of precursor programmes with isotopically- and chemically-tailored steels and helium ion implantation. Fundamental modelling is important for understanding the extrapolation of these results to a true fusion neutron irradiation condition, and has many other key roles, identified in the Roadmap's recommendations.

**Acknowledgments**

The authors gratefully acknowledge the support of their home institutions and research funders in this work, which was carried out within the framework of the European Fusion

**Figure Captions**

*Figure 1:* Flow chart showing schematically the process of risk analysis and mitigation to select Baseline and Risk Mitigation Materials in a particular category.

*Figure 2:* Temperature variation of yield strength of laboratory heats of US HT RAFM steels (1537-1539, and Mod-NF616), compared to US developed ODS steel (PM2000) and US and Japanese RAFM steels (NF616 and F82H) (data from [54])

*Figure 3(a):* Comparative TEM scans of nano-precipitate density in normally treated NF616 steel and the same alloy subjected to hot rolling ('TMT') during plate manufacture (taken from ref [60])

*3(b):* Comparative photo-micrographs of prior austenite grain size for the NF616 steel under the manufacture conditions as in 3(a) (taken from ref [60]).

*Figure 4(a):* Larson-Miller Parameter (LMP) plotted for various FM steels. The data points are for the AHT FM steels from ref [64] (arrows indicate that the specimens\ did not fail at the plotted position). The curves are representative of the databases for: type 92 (9Cr, 0.5Mo, 2W) commercial FM steel – source ECCC2005;EUROFER and 9Cr ODS (taken from ref [65])

*4(b:)* LMP re-plotted for the AHT FM steels in figure 4(a), with LMP values corresponding to 5 years at temperature and the indicated stress.

*Figure 5:* Effect of fission neutron irradiation on the yield strength of RAFM steels irradiated at 300°C, compared to low-sink density reactor pressure vessel (RPV) steel, showing the delayed onset of radiation hardening. Adapted from ref [69].

*Figure 6:* Damage levels (dpa/fpy) for tungsten (left hand graph) and copper alloy (right hand graph) for locations in the divertor region (middle graph) of the 1.8GW$_{th}$ DEMO Roadmap concept. Strike zones are shown ringed on the dpa graphs, indicating the shielding effect of the strike zone channels. Data are scaled from ref [14]

*Figure 7:* Effective sputtering yields for tungsten as a function of temperature for different species, assuming $T_e=T_i$. The black curve gives the total yield for the species mix indicated. The right picture shows the corresponding tungsten influx for the left graph conditions. A 1mm/fpy DEMO divertor erosion limit is shown corresponding to divertor target plasma temperature $\leq 5eV$.

*Figure 8:* Cartoon schematic of a cross-section through a water-cooled divertor with tungsten and CuCrZr composition, showing relevant temperature limitations to avoid various performance-degrading effects.

*Figure 9(a):* Comparative time histories of dose-rate of the outboard mid-plane of a DEMO vacuum vessel constructed of 316LN (ITER Grade) subjected to a 30 fpy lifetime, and the outboard mid-plane EUROFER First Wall of the Helium-cooled Pebble Bed (HCPB) blanket, subjected to a 5 fpy lifetime. The results are for a simulation of a 2.7 GW$_{th}$ DEMO reactor, described in references [14-16]. The 'ITER Administrative Limit' indicated refers to the 100 μSv.hr$^{-1}$ limit for items available for hands-on maintenance after 12 hours post pulsing shutdown, as defined in the ITER Safety Case [89].

*(b):* Comparative time histories of specific or total activiation of the same outboard mid-plane components' materials – conditions as listed in (a).

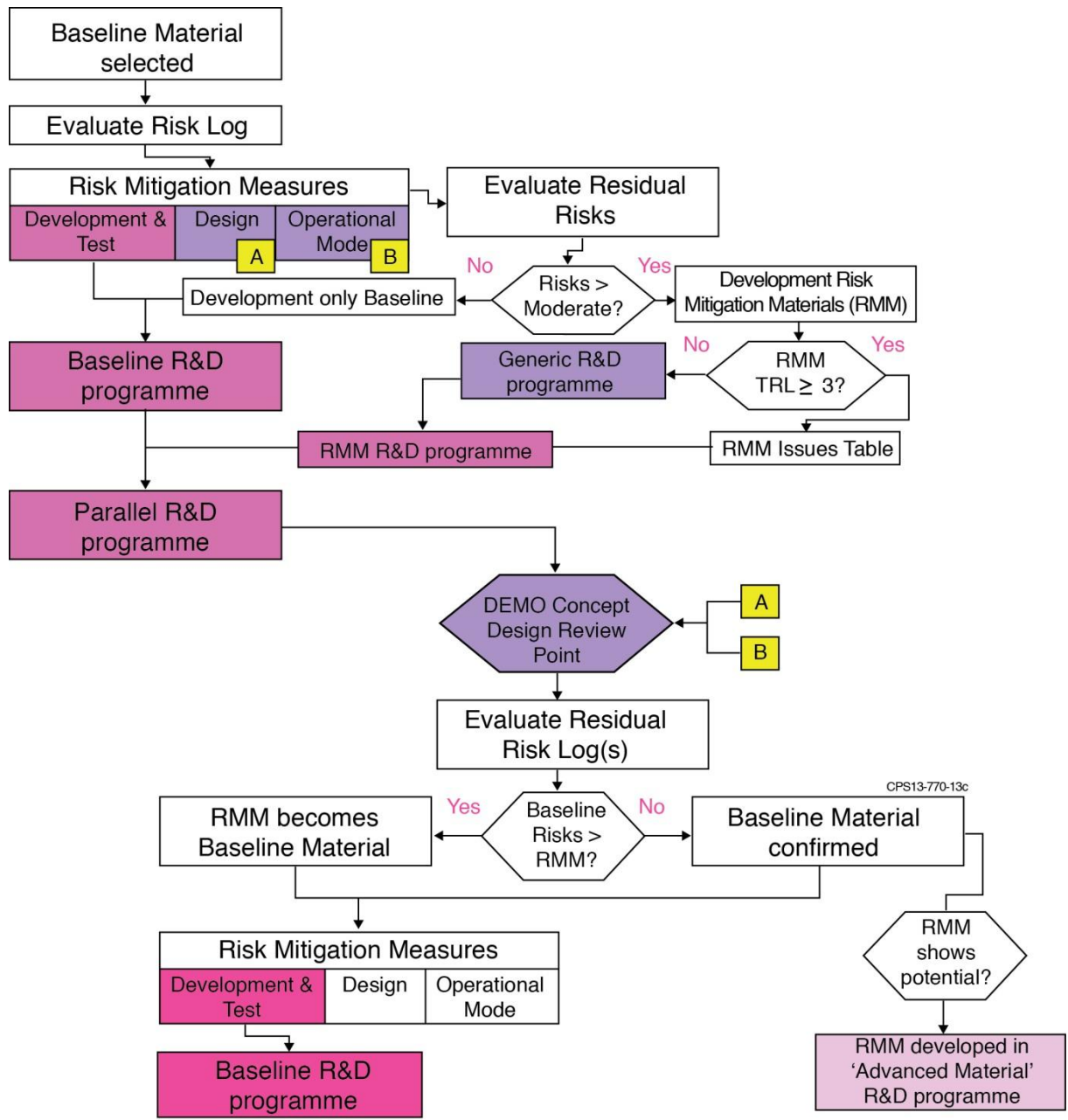

*Figure 1*

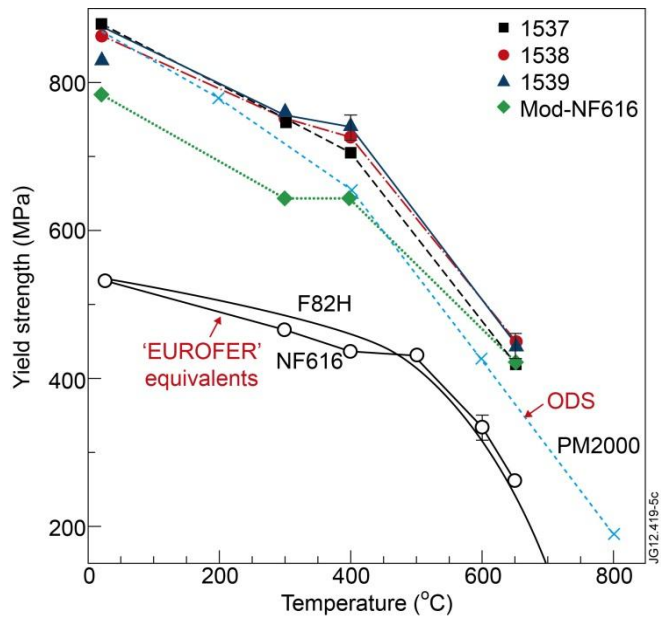

*Figure 2*

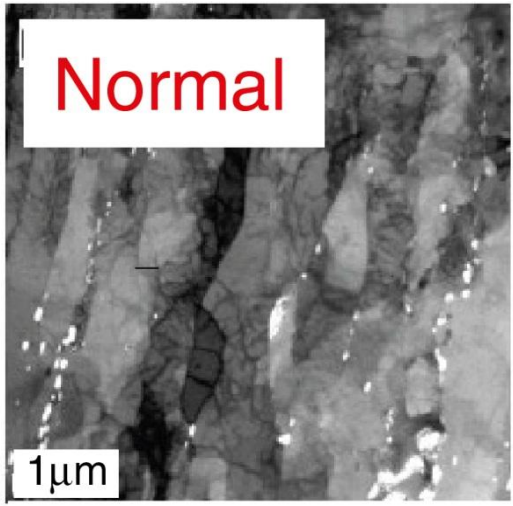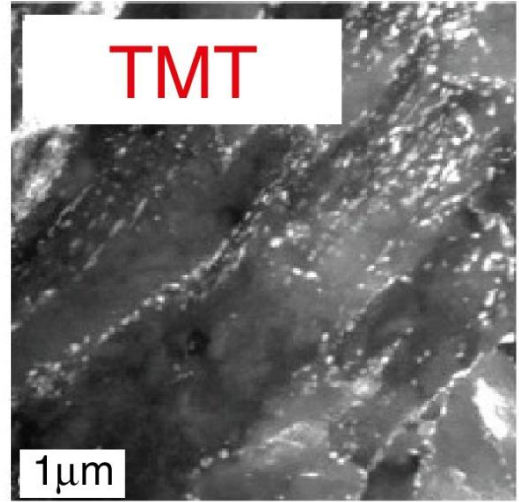

*Figure 3(a)*

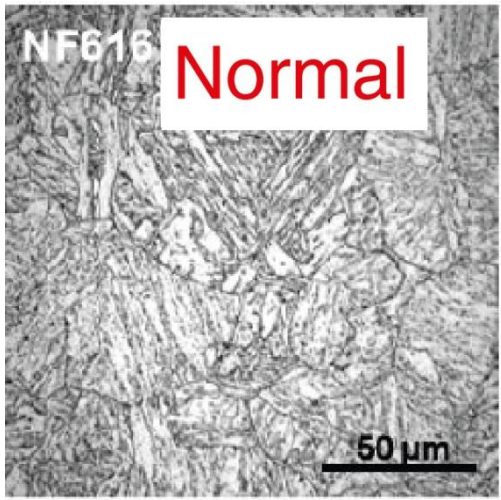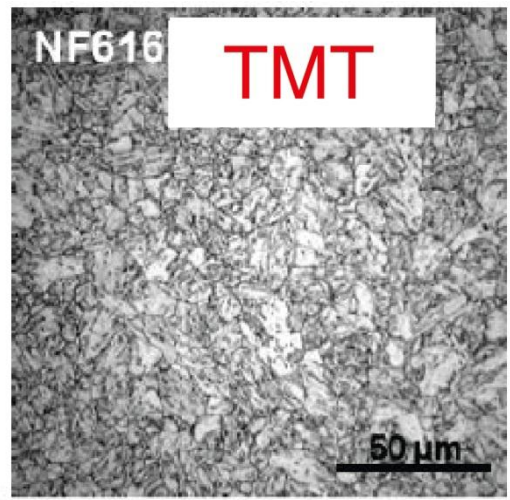

*Figure 3(b)*

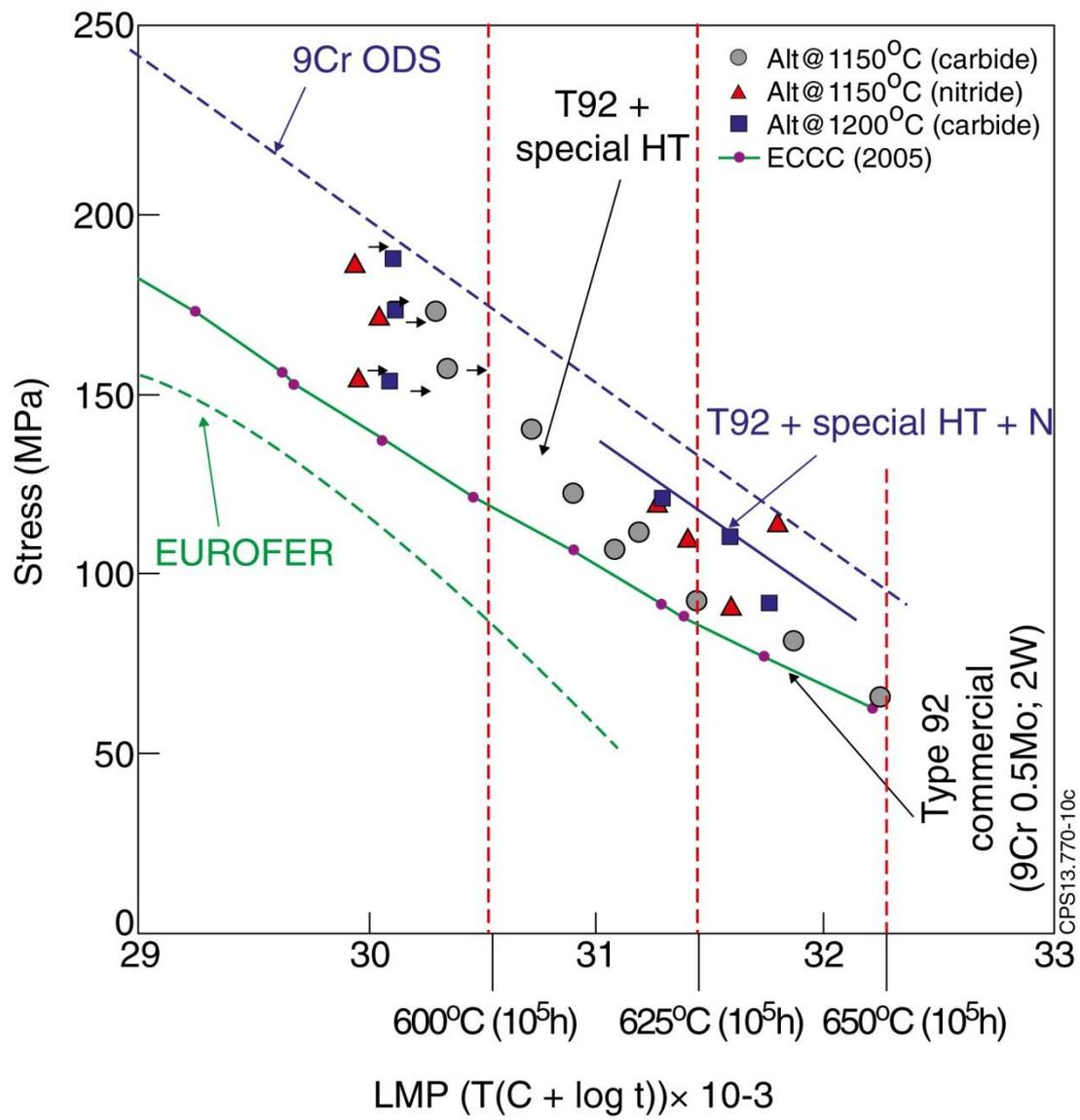

*Figure 4(a)*

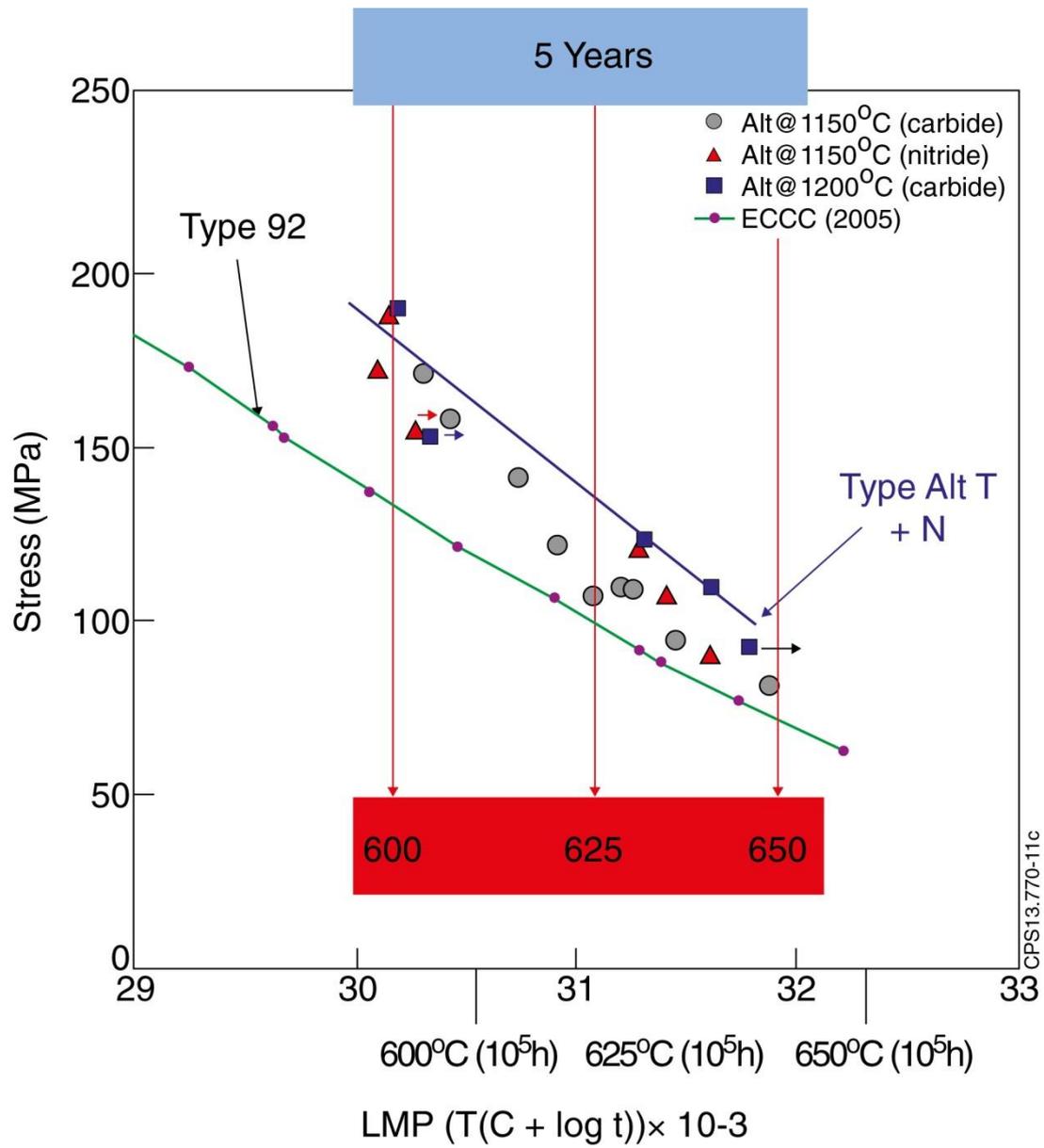

*Figure 4(b)*

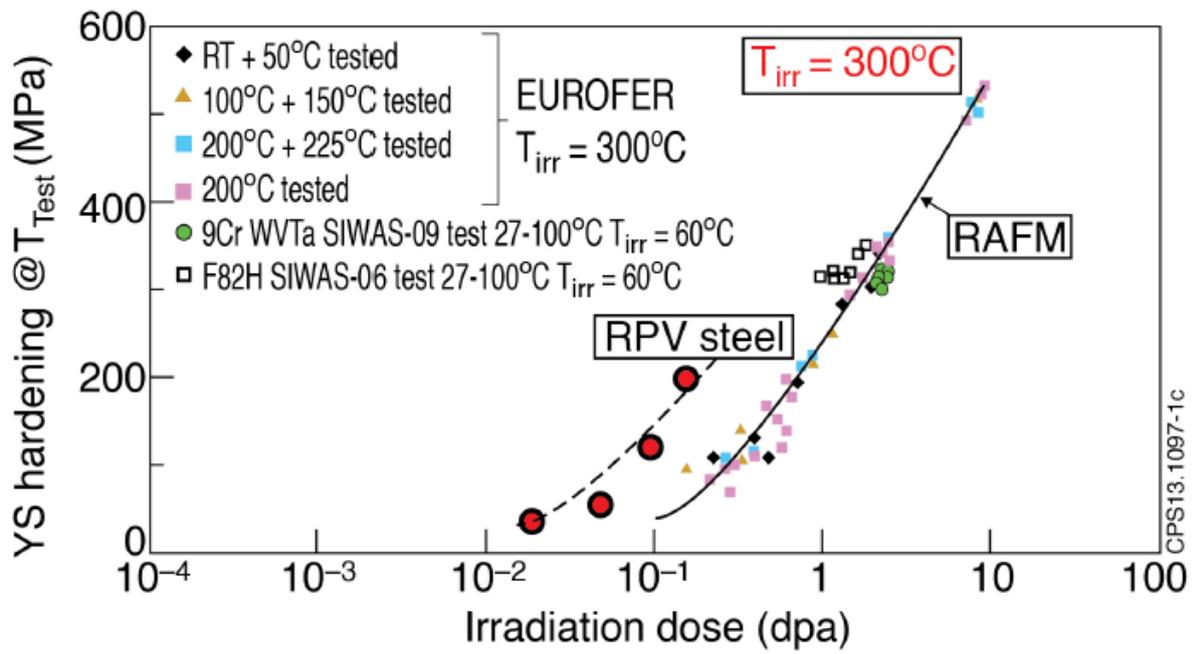

*Figure 5*

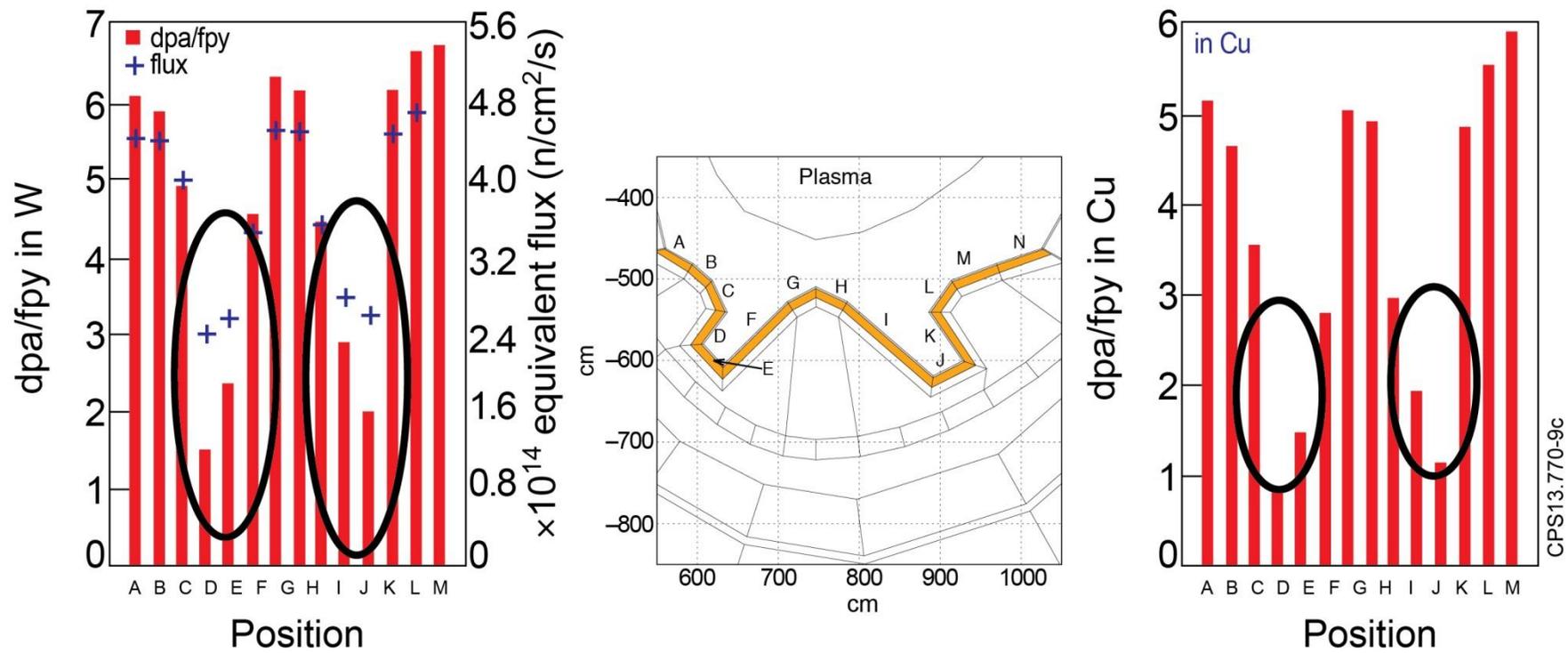

*Figure 6*

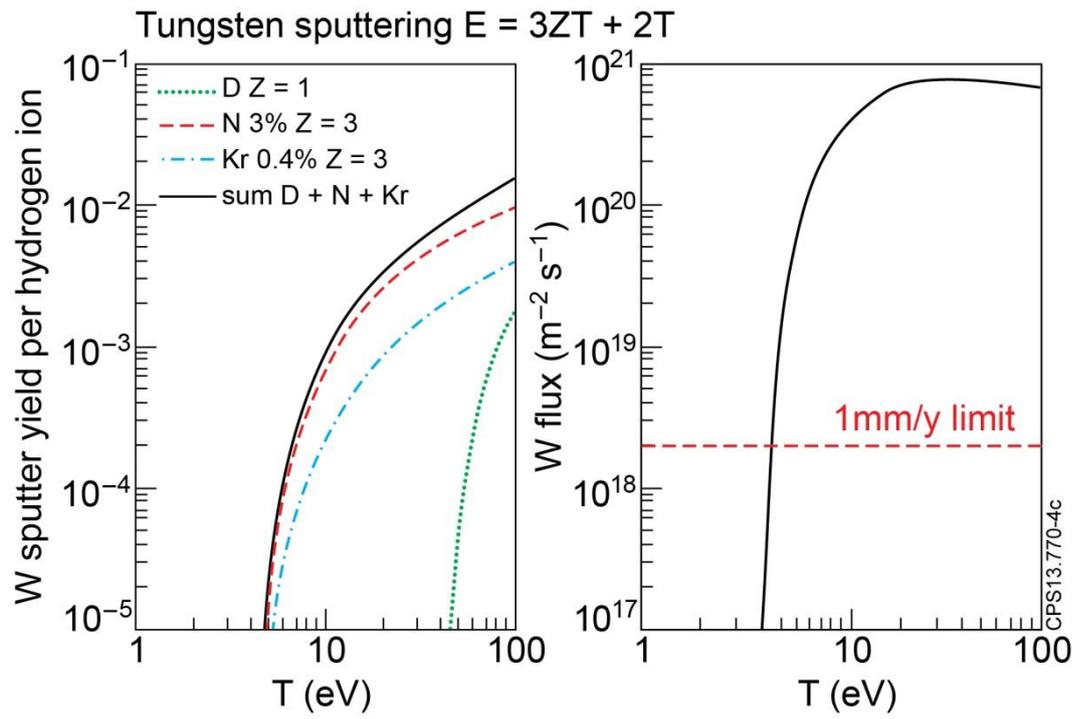

*Figure 7*

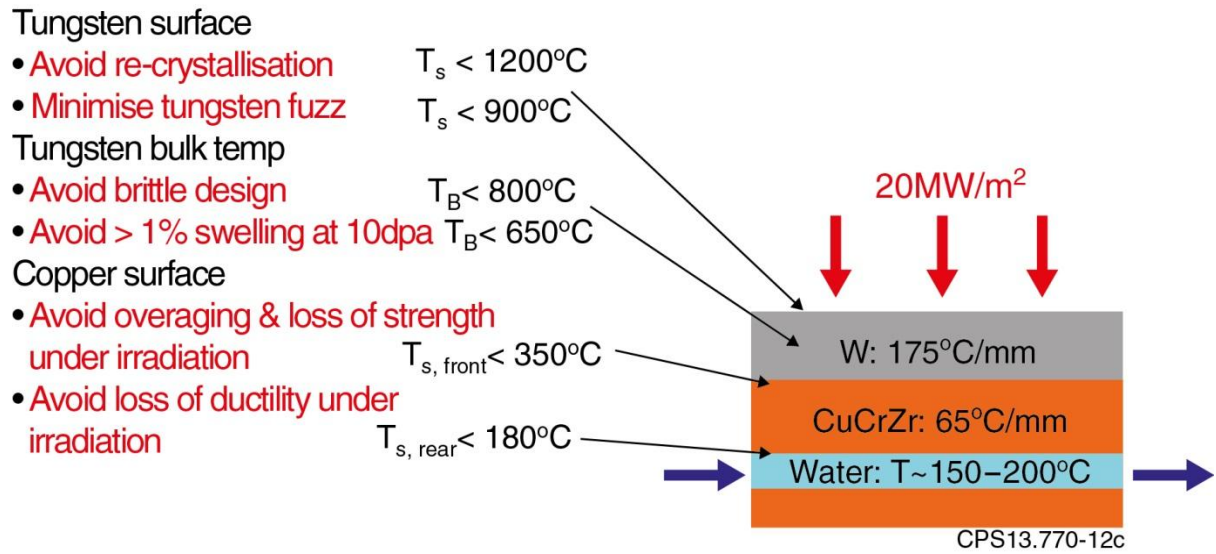

*Figure 8*

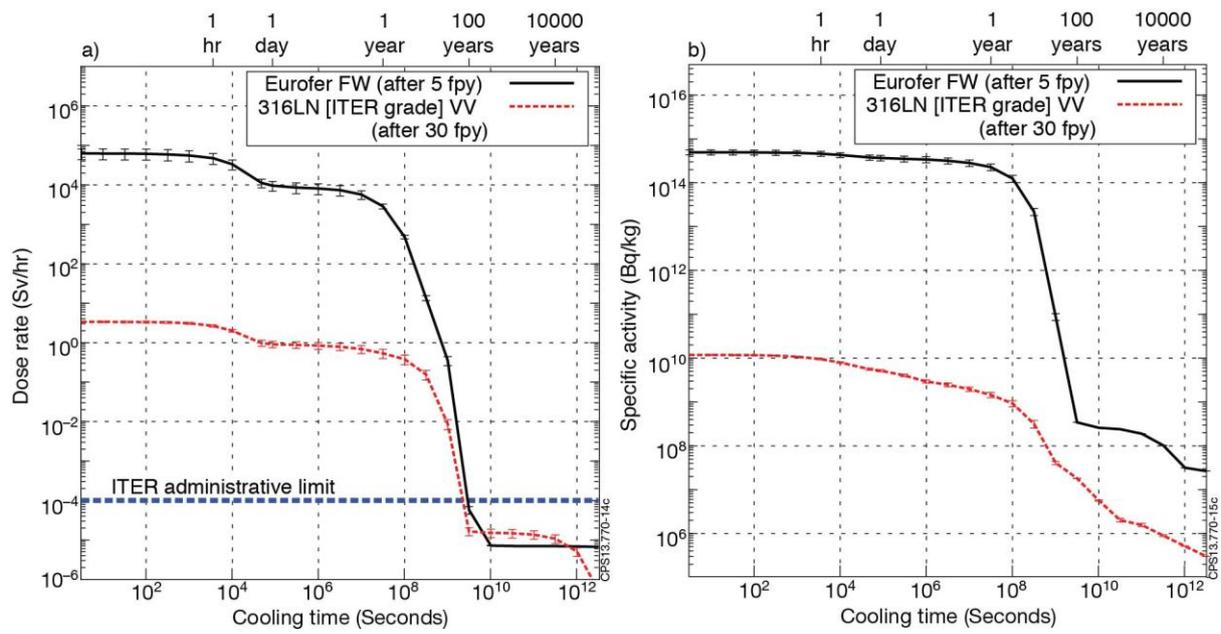

*Figure 9*